\RequirePackage{lineno}
\documentclass[aps,twocolumn,showpacs,preprintnumbers,amsmath,amssymb]{revtex4}
\usepackage{graphicx}
\usepackage{dcolumn}
\usepackage{epstopdf}
\usepackage{color}
\usepackage{amssymb}
\usepackage{bm}
\usepackage{subfigure}
\definecolor{dgreen}{cmyk}{1.,0.,1.,0.2}        
\definecolor{orange}{cmyk}{0.,0.353,1.,0.}    

\def\bea {\begin{eqnarray}}
\def\eea {\end{eqnarray}}

\def\be {\begin{equation}}
\def\ee {\end{equation}}

\usepackage{hyperref}
\RequirePackage{lineno}

\def\bea {\begin{eqnarray}}
\def\eea {\end{eqnarray}}

\def\be {\begin{equation}}
\def\ee {\end{equation}}

\newcommand{\Eta}{$\eta$}

\newcommand{\aveNpart}{$\langle N_{\rm part}\rangle$}
\newcommand{\sNN}{$\sqrtsign {s_{\rm NN}}$}

\newcommand{\sgg}{$\sigma_{\rm gg}$}
\newcommand{\Ebj}{$\epsilon_{\rm Bj}\,\tau$}

\begin{document}

\title{Beam energy dependence of pseudorapidity distributions of
  charged particles produced in heavy-ion collisions at RHIC and LHC energies}

\author{Sumit Basu}\affiliation{Variable Energy Cyclotron Centre, Kolkata 700064, India}
\author{Tapan K. Nayak}\affiliation{Variable Energy Cyclotron Centre, Kolkata 700064, India}
\author{Kaustuv Datta}\affiliation{Department of Physics, Reed College, Portland, OR 97202, USA
}

\date{ \today}

\begin{abstract}

Heavy-ion collisions at the Relativistic Heavy Ion Collider at 
Brookhaven National Laboratory and the Large Hadron Collider at CERN 
probe matter at extreme conditions of temperature and energy density. 
Most of the global properties of the collisions can be extracted from the 
measurements of charged particle multiplicity and pseudorapidity~($\eta$)
distributions. We have shown that the available experimental data on
beam energy and centrality dependence of \Eta-distributions in heavy-ion
(Au+Au or Pb+Pb) 
collisions from \sNN~=~7.7~GeV to 2.76~TeV 
are reasonably well described by the AMPT
model, which is used for further exploration.  
The nature of the \Eta-distributions has
been described by a double Gaussian function
using a set of fit parameters, which exhibit a regular pattern as a
function of beam energy. By extrapolating the
parameters to a higher energy of \sNN~=~5.02~TeV, we have obtained
the charged particle multiplicity
densities, \Eta-distributions and energy densities for
various centralities. 
Incidentally, these results match well with some of
the recently published data by the ALICE collaboration.
\end{abstract}

\maketitle

\section{Introduction}

The primary goal of colliding heavy-ions at 
ultra-relativistic energies 
is to study nuclear matter under extreme conditions, in which hadronic 
matter is expected to undergo a phase transition to a new state of 
matter, the Quark-Gluon Plasma (QGP)~\cite{intro1,intro2}. 
Quantum Chromodynamics (QCD), the theory of strong interactions, suggests that 
at high temperatures and energy densities, 
nuclear matter melts down to this new phase of deconfined quarks and gluons. 
Recent Lattice QCD 
calculations~\cite{lattice1,lattice2} 
indicate that transition from hadronic matter to 
QGP occurs at a critical temperature of $T_C \sim 155$ MeV and 
critical energy density of $\epsilon_C \sim 0.7-1.9$~GeV/fm$^3$. 
The QGP research programs at the Relativistic Heavy Ion Collider (RHIC) at 
Brookhaven National Laboratory and the Large Hadron Collider (LHC) at 
CERN are on a quest to unearth the physics of deconfinement and 
vacuum, and to understand how matter behaved within a few microseconds 
after the birth of our Universe. 
With the first phase of the beam energy scan program at RHIC during 2010 
and 2011, data for Au+Au collisions at 
a nucleon-nucleon (NN) centre-of-mass energy (\sNN)~from 7.7~GeV to 200~GeV are 
available. The main aim of this program is to probe the onset 
of deconfinement and to locate the QCD Critical Point~\cite{scan}. The LHC has 
collided Pb+Pb beams at \sNN~=~2.76~TeV during the first phase of its 
operation (2010 and 2011). During the first year of the second phase 
of LHC operation in 2015,
data for Pb+Pb collisions at \sNN~=~5.02~TeV are collected. Thus with 
the combination of RHIC and LHC, high quality data for heavy-ion 
collisions have now been available over quite a broad energy range. At 
the same time a large number of models have emerged which 
attempt to analyze and explain the data and extract physical parameters 
\cite{rainer,ugo,chinese, stephans, shao}.

Global observables such as charged particle multiplicity
distributions, pseudorapidity (\Eta) distributions, momentum spectra,
particle ratios, 
size of the fireball, and azimuthal anisotropy provide majority of the
valuable information for thermal and chemical analysis of the
freeze-out conditions~\cite{raghu1,raghu2}.  
The \Eta-distribution of charged particles is one of the
most basic and most important observables to characterize the colliding 
system and to understand the phase transition. All the observables in
heavy-ion collisions scale with the number of particles. So the knowledge
of the particle density is essential for validating any measurement.
The pseudorapidity particle
density at mid-rapidity, along with transverse energy per particle
provides the energy density of the fireball using the Bjorken estimation~\cite{bjorken}. 
The pseudorapidity distributions
are intimately connected to the energy density of the emitting source
and provide an important test-bed for validating
theoretical models, which attempt to describe the conditions in the
early phases of the collision.

Experimental data for \Eta-distributions have been reported
for all the collider energies at RHIC~\cite{phobos,star_et2}
and LHC~\cite{alice2,alice2-2,cms,atlas1,atlas2}. In this article, we make a compilation of
some of the available data in terms of the 
variation of pseudorapidity distributions of charged particles with beam energy and
collision centrality. We make a similar study using the string melting
mode of the A Multi-Phase Transport (AMPT) model and make a comparison with the available data.
In this model, different values of parton
cross sections are used to explain the data at LHC.
The pseudorapidity distributions, 
both from data and the AMPT model, of charged particles from \sNN~=7.7~GeV to 2.76~TeV
are fitted by a double Gaussian
function. These parameters show interesting
trends as a function of beam energy. 
Extrapolating the parameters to higher energies, we obtain the 
\Eta-distribution for \sNN~=~5.02~TeV. It is observed that the
pseudorapidity density at mid-rapidity matches well with the recently
reported data from ALICE~\cite{alice5}. 
Furthermore, we extract the value of initial 
energy density for collisions at \sNN~=~5.02~TeV.

The paper is organized as follows. In Section~II, we discuss the AMPT
model which is used to compare the data results. In Section~III, we
present the compilation of pseudorapidity distributions for data and
AMPT. In Section~IV, we make an analysis of the shapes of the
pseudorapidity distributions and present the results of the fit
parameters. Energy dependence of charged particle multiplicity
densities, pseudorapidity distributions and energy densities are
presented. We conclude the paper with a summary in Section~VI.

\section{AMPT settings}

The AMPT model~\cite{ampt1} provides a framework to study relativistic heavy-ion
collisions. It incorporates essential stages of heavy ion collisions
from the initial condition to final observables on an event-by-event
basis, including the parton cascade, hadronization and the 
hadron cascade~\cite{ampt2,ampt3,ampt4}. 
The model can generate events in two different modes: (a) default, and
(b) string melting (SM).
Initial conditions for both the modes are taken from
HIJING~\cite{hijing}, where
two Wood-Saxon type radial density profile are taken for colliding nuclei.
The multiple scattering among the nucleons of two heavy ion nuclei
are governed by the eikonal formalism. The particle production has two distinct
sources, from hard and soft processes, depending on the momentum transfer
among partons. 
In the default mode, energetic partons cascade through Zhang's Parton
Cascade (ZPC) before the strings 
and partons are recombined and the strings are fragmented via the Lund string fragmentation function, 
\begin{equation}
f(z) \propto~ z^{-1}(1-z)^{a}exp(-b m^2_{T}/z),
\end{equation}
where a and b are the Lund string fragmentation function parameters, taken to be
0.2 and 2.2. ART (A Relativistic Transport model for hadrons)~\cite{art} is used to describe
how the produced hadrons 
will interact.
In the String Melting mode, the strings produced from HIJING are
decomposed into partons which are fed into the parton cascade along
with the minijet partons. The partonic matter is then turned into
hadrons through the coalescence model~\cite{coal1,coal2} and the hadronic interactions are
subsequently modeled using ART.
The Default mode describes the evolution of collision in
terms of strings and minijets followed by string fragmentation,
and the
String Melting mode includes a fully partonic QGP phase that
hadronizes through quark coalescence. 

In both the modes of AMPT, 
Boltzmann equations are solved using ZPC with total parton elastic scattering cross section, 
\begin{equation}
\sigma_{gg} = \frac{9\pi \alpha^2_s}{2\mu^2} \frac{1}{1+\mu^2/s} \approx~\frac{9\pi \alpha^2_s}{2\mu^2},
\end{equation}
where $\alpha_s$ is the strong coupling constant, $s, t$ are the
Mandelstam variables and $\mu$ is the Debye screening mass. 
Here, $\alpha_s$ and $\mu$ are the key deciding factors for multiplicity
yield at a particular centrality of given energy, 
and they are taken as 0.47 and 3.22, corresponding to \sgg~=~10~mb. For a beam
energy range 7.7 GeV to 2.76~TeV we found  global observables like
pseudorapidity density~\cite{alice2}, transverse momentum
distribution~\cite{dronika}, particle ratio~\cite{ampt1}, higher
harmonic anisotropic flow~\cite{dronika} like v2, v3 are 
within the range of experimental error. 
We have carried out a comparison study for
different observables by varying $a,b, \alpha_s$ and $\mu$ corresponding to 1.5~mb, 3~mb, 6~mb
and 10~mb cross sections. The model
therefore provides a convenient way to investigate expectations for a
variety of observables with and without a QGP phase.

\section{Pseudorapidity Distributions - Data and AMPT}

\begin{figure}[tbp]
\centering \includegraphics[width=0.49\textwidth]{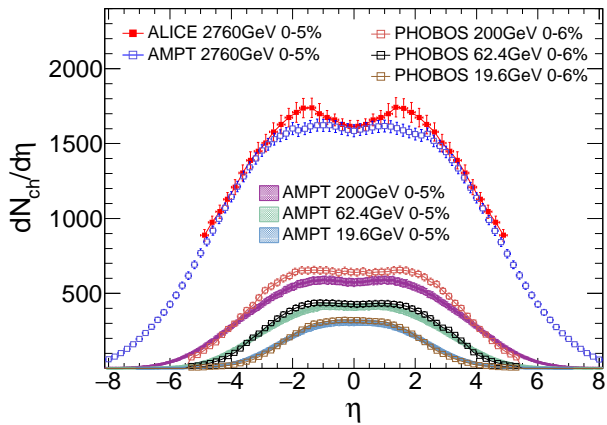} 
\caption{Beam energy dependence of charged particle pseudorapidity
  distributions. Results from PHOBOS~\cite{phobos} and ALICE~\cite{alice2,alice2-2} for
  central collisions are shown along with calculations from
  the string melting mode of AMPT model. 
}
\label{rap1}
\end{figure}

Pseudorapidity distributions of charged particles have been reported by
fixed target as well as collider experiments. In this article, we
concentrate on the results of collider experiments at RHIC and LHC.
In Fig.~\ref{rap1}, we present the experimental results from the
PHOBOS experiment~\cite{phobos} at RHIC for central 
Au+Au collisions at \sNN~=~19.6, 62.4 and
200~GeV, and from the
ALICE experiment~\cite{alice2} at LHC for 
Pb+Pb collisions at \sNN~=~2.76~TeV. It is observed that the
distributions are symmetric around the mid-rapidity as they should
be, but the dip structure at \Eta~=~0 gets more prominent with the
increase of collision energy. For the LHC energy, the dip increases in
going from peripheral to central collisions. The magnitude of the dip
depends on the particle composition of the charged particles as the
dip is more prominent for heavier particles like protons and
anti-protons compared to pions. 

In the present study, we have generated AMPT events with SM mode for different
collision energies and collision centralities. 
The total parton elastic scattering cross section from 7.7 GeV to 200~GeV 
at RHIC energies is taken as \sgg~=~10~mb and for 2.76~TeV at LHC energy, it is 
chosen to be 1.5~mb. It is observed that with these settings AMPT can
describe the data for 
transverse momentum spectra and flow~\cite{dronika}.
The results of AMPT model calculations for \Eta-distributions are
superimposed on Fig.~\ref{rap1}. The AMPT results describe the
data at RHIC energy well. For \sNN~=~2.76~TeV, 
the data at mid-rapidity are well
described by AMPT, but discrepancies are observed at other
\Eta-ranges especially at the peaks.  

\begin{figure}[tbp]
\centering \includegraphics[width=0.49\textwidth]{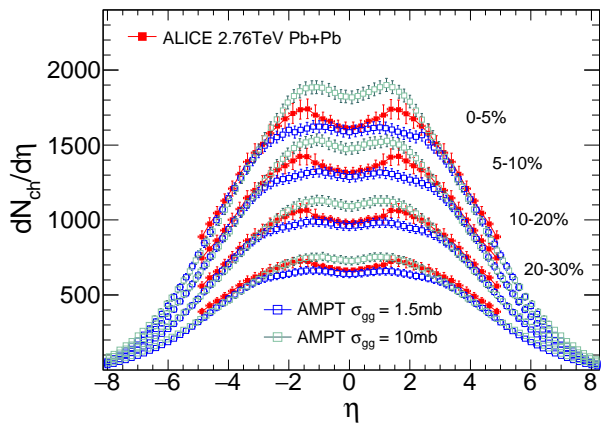} 
\caption{Centrality dependence of charged particle pseudorapidity distributions for Pb+Pb collisions at
\sNN~=~2.76~TeV with the data from ALICE experiment~\cite{alice2,alice2-2} and 
those from the AMPT model for two settings of total parton scattering 
cross section~(\sgg).}
\label{rap2}
\end{figure}

In Fig.~\ref{rap2}, \Eta-distributions for LHC data at
\sNN~=~2.76~TeV for four centralities along with AMPT model data for 
two different parton scattering cross sections (1.5~mb and 10~mb) are shown.
It is seen that the AMPT results with 1.5~mb matches the
mid-rapidity value quite well. The distributions with 10~mb, match
the shape of the data distribution very well, but miss the value at
mid-rapidity. Henceforth, parton cross sections are kept at 1.5~mb for all
calculations at LHC energies.

\section{Shapes of pseudorapidity distributions}

Further studies have been performed to investigate the centrality-wise
variation of shape of the \Eta-distributions for heavy-ion collisions, ranging
from 7.7~GeV to 2.76~TeV. 
For central Au+Au collisions at RHIC energies, the distributions has been
fitted by~\cite{phenix_et1}:
\begin{equation}
\frac{dN_{\rm ch}}{d\eta} = \frac{c\sqrt{1-1/(\alpha \cosh \eta)^2}}{1+e^{(|\eta|-\beta)/a}},
\label{WS_fit}
\end{equation}
where $a, c, \alpha,$ and  $\beta$ are fit parameters. 

Figures~\ref{rap1} and \ref{rap2} show that
the \Eta-distributions exhibit double 
Gaussian nature, both for experimental data and AMPT. This double Gaussian nature
is more prominent for higher collision energies and central collisions.
The shapes can be represented by
double Gaussian distributions of the form,
\begin{eqnarray}
A_{\mathrm{1}} e^{-({\eta_1^2}/{2\sigma_{\mathrm{1}}^2})} - 
A_{\mathrm{2}} e^{-({\eta_2^2}/{2\sigma_{\mathrm{2}}^2})},
\end{eqnarray}
where the fit parameters, $A_1, A_2$ are the amplitudes, $\eta_1, \eta_2$ are the peak
positions, and $\sigma_1, \sigma_2$ are the widths of the two Gaussian
distributions. The fit parameters represent the shapes of the distribution.

\begin{figure*}[tbp]
\centering \includegraphics[width=0.91\textwidth]{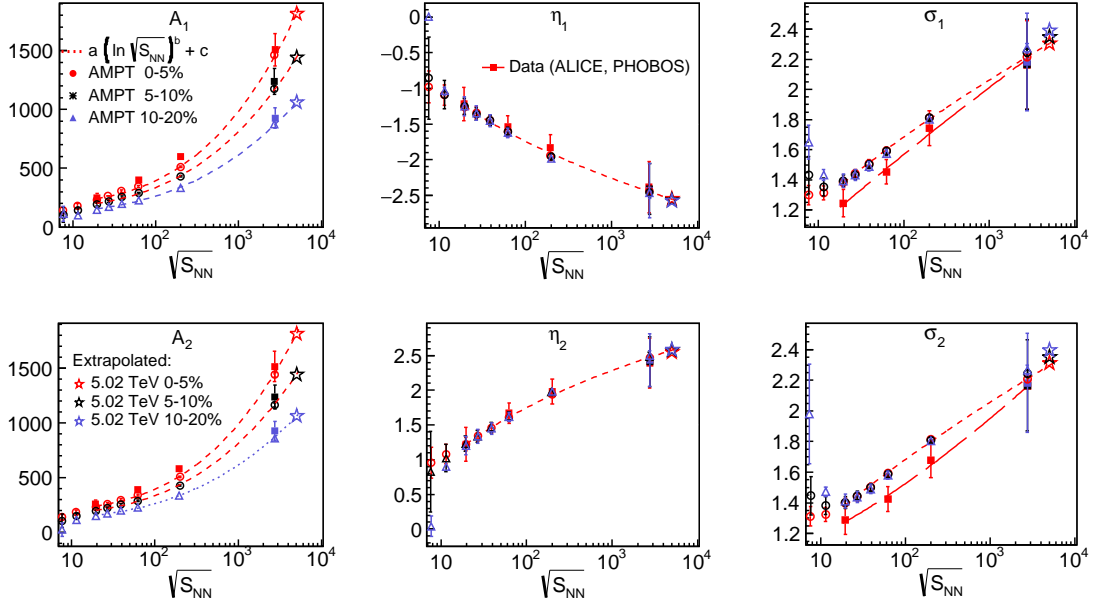} 
\caption{Fit parameters of the double Gaussian fit to the 
  \Eta-distributions obtained from the AMPT model for Au+Au collisions 
  from \sNN~=~7.7~GeV to 200~GeV and Pb+Pb collision at \sNN~=~2.76~TeV. 
  Extrapolated values of the parameters for \sNN~=~5.02~TeV are 
  also plotted in the figures.}
\label{param}
\end{figure*}
{\begin{table*}
	\caption{Parameters of Double Gaussian fits to the 
		\Eta-distributions of Au+Au 
		collisions from \sNN~=~7.7~GeV to 200 GeV, and Pb+Pb 
		collisions at 2.76~TeV.  Extrapolated parameters for 
		\sNN~=~5.02~TeV are presented. 
	}
	\begin{center}
		\begin{tabular}{ |c|c|c|c|c|c|c|c| }
			\hline 
			\sNN (GeV) & Centrality  (\%)   & $A_1 $ & $\eta_{1} $ & $\sigma_{1}$  & $A_2$ & $\eta_{2}$ & $\sigma_{2}$ \\ \hline 
			{   7.7} & 0-5 & $134.93\pm25.67$ &  $-0.987\pm0.223$ & $1.294\pm0.063$ & $139.120\pm25.05$ & $0.225\pm0.225$ & $1.312\pm0.064$\\ 
			& 5-10 & $102.46\pm63.61$ & $-0.862\pm0.576$ & $1.432\pm0.124$ & $106.84\pm62.43$ & $0.825\pm0.581$ & $1.446\pm0.126$\\ 
			& 10-20 & $112.36\pm61.54$ & $-0.004\pm0.047$ & $1.648\pm0.112$ & $26.63\pm61.93$ & $0.042\pm0.15$ & $1.980\pm0.325$\\ \hline 
			{   11.5} & 0-5 & $178.72\pm17.98$ & $-1.097\pm0.142$ & $1.314\pm0.045$ & $180.78\pm17.72$ & $0.143\pm0.143$ & $1.323\pm0.059$\\ 
			& 5-10 & $142.25\pm22.25$ & $-1.091\pm0.199$ & $1.354\pm0.059$ & $150.42\pm21.37$ & $1.016\pm0.199$ & $1.380\pm0.059$\\ 
			& 10-20 &  $100.56\pm2.19$ & $-1.037\pm0.051$ & $1.433\pm0.035$ & $114.22\pm2.89$ & $0.892\pm0.037$ & $1.473\pm0.030$\\ \hline 
			{   19.6} & 0-5 & $226.70\pm13.19$  & $-1.269\pm0.098$ & $1.383\pm0.034$ & $232.85\pm12.90$ & $1.223\pm0.098$ & $1.399\pm0.034$\\ 
			& 5-10 &  $190.92\pm12.63$ & $-1.255\pm0.111$ & $1.392\pm0.038$ & $194.42\pm12.44$ & $1.224\pm0.111$ & $1.402\pm0.038$\\ 
			& 10-20 & $147.38\pm11.90$ & $-1.254\pm0.132$ & $1.393\pm0.045$ & $151.82\pm11.58$ & $1.203\pm0.132$ & $1.411\pm0.045$\\ \hline 
			{   27} & 0-5 & $260.45\pm11.09$ & $-1.344\pm0.082$ & $1.441\pm0.029$ & $260.59\pm11.11$ & $1.345\pm0.082$ & $1.441\pm0.029$\\ 
			& 5-10 &  $218.19\pm10.28$ & $-1.361\pm0.089$ & $1.433\pm0.032$ & $221.92\pm10.10$ & $1.326\pm0.089$ & $1.446\pm0.032$\\ 
			& 10-20 & $171.77\pm9.06$ & $-1.346\pm0.101$ & $1.432\pm0.036$ & $172.96\pm8.99$ & $1.333\pm0.100$ & $1.437\pm0.036$\\ \hline 
			{   39} & 0-5 & $299.95\pm9.79$ & $-1.444\pm0.069$ & $1.508\pm0.026$ & $297.48\pm9.87$ & $1.457\pm0.070$ & $1.502\pm0.026$\\ 
			& 5-10 &  $254.58\pm8.83$ & $-1.450\pm0.074$ & $1.501\pm0.027$ & $253.57\pm8.87$ & $1.455\pm0.074$ & $1.499\pm0.027$\\ 
			& 10-20 & $199.13\pm7.64$ & $-1.455\pm0.082$ & $1.490\pm0.030$ & $199.39\pm7.62$ & $1.450\pm0.082$ & $1.490\pm0.031$\\ \hline 
			{   62.4} & 0-5 & $341.36\pm8.04$ & $-1.605\pm0.057$ & $1.595\pm0.022$ & $340.53\pm8.07$ & $1.670\pm0.057$ & $1.594\pm0.022$\\ 
			& 5-10 &  $288.93\pm7.14$ & $-1.608\pm0.061$ & $1.589\pm0.023$ & $287.59\pm7.16$ & $1.619\pm0.061$ & $1.587\pm0.023$ \\ 
			& 10-20 & $225.61\pm6.07$ & $-1.625\pm0.066$ & $1.576\pm0.026$ & $225.71\pm6.035$ & $1.615\pm0.067$ & $1.580\pm0.026$\\ \hline 
			{   200} & 0-5 & $507.18\pm6.81$ & $-1.947\pm0.041$ & $1.812\pm0.016$ & $506.93\pm6.77$ & $1.940\pm0.041$ & $1.816\pm0.016$\\ 
			& 5-10 &  $430.61\pm5.97$ & $-1.958\pm0.043$ & $1.813\pm0.017$ & $429.01\pm5.99$ & $1.965\pm0.043$ & $1.809\pm0.017$\\ 
			& 10-20 & $334.48\pm4.97$  & $-1.982\pm0.047$ & $1.804\pm0.019$ & $334.30\pm4.97$ & $1.979\pm0.047$ & $1.803\pm0.019$\\ \hline 
			{   2760} & 0-5 & $1458.69\pm19.63$ & $-2.442\pm0.054$ & $2.215\pm0.022$ & $1439.93\pm19.75$ & $2.471\pm0.054$ & $2.207\pm0.022$\\ 
			& 5-10 &  $1174.33\pm18.66$ & $-2.462\pm0.063$ & $2.245\pm0.026$ & $1159.96\pm18.68$ & $2.475\pm0.064$ & $2.244\pm0.026$\\ 
			& 10-20 & $872.77\pm16.66$ & $-2.465\pm0.075$ & $2.274\pm0.031$ & $859.07\pm16.63$ & $2.493\pm0.076$ & $2.266\pm0.032$\\ \hline 
			{   5020 } & 0-5 &  $1814.52\pm27.92$ & $-2.554\pm0.082$ & $2.304\pm0.039$ & $1815.19\pm28.00$ & $2.549\pm0.085$ & $2.311\pm0.040$\\ 
			(extrapolated) & 5-10 &  $1441.13\pm21.23$ & $-2.573\pm0.091$ & $2.348\pm0.046$ & $1442.23\pm22.80$ & $2.579\pm0.092$ & $2.352\pm 0.050$\\ 
			& 10-20 & $1059.63\pm17.81$ & $-2.575\pm0.107$ &
                                                                    $ 2.389\pm0.051$
                                                                                               & $1061.73\pm18.10$ & $2.585\pm0.110$ & $2.395\pm0.550$  \\ \hline 
		\end{tabular}
	\end{center}
	\label{table1}
\end{table*}}

Both the experimental data and AMPT distributions are 
fitted with the double Gaussian functional form as above and the fit parameters are 
extracted. 
The fit parameters are presented in Fig.~\ref{param}
as a function of collision energy for experimental data and AMPT
calculations. All the errors shown in the this figure correspond to
the error in fitting.
The Gaussian fit parameters follow the following trends: \\
\noindent 
(i) The normalization parameters, $A_{\rm 1}$ and $A_{\rm 2}$ 
increase with the increase of
beam energy as per expectation. These parameters for available experimental data
and AMPT are observed to be close together. \\
\noindent 
(ii) The values of $\eta_{\rm 1}$ and 
$\eta_{\rm  2}$ represent the peak positions in the
\Eta~distribution. As expected, $\eta_{\rm 1}$ and $\eta_{\rm 2}$
show opposite trends with the increase of the beam
energy. This means that the peak positions in $\eta$
spread out more with the increase of beam energy.
It is to note that the values of $\eta_{\rm 1}$ and $\eta_{\rm 2}$ for
data and AMPT are close together. \\
\noindent
(iii) The widths ($\sigma_1$ and $\sigma_2$),
of the \Eta-distributions increase as a function
of beam energy. For lower collision energies, the widths extracted
from data are smaller than those of AMPT, but are close together at
higher energies. 

\begin{figure}[tbp]
\centering \includegraphics[width=0.49\textwidth]{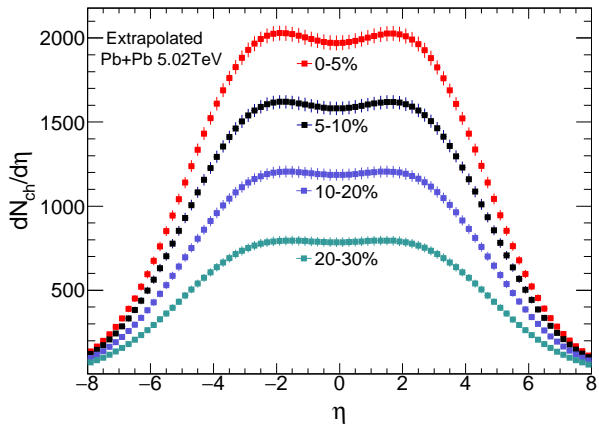} 
\caption{\Eta-distributions for Pb+Pb collisions at \sNN~=~5.02~TeV
  for different centralities. The distributions are obtained from the
  extrapolated AMPT parameters from lower energies.}
\label{eta5020}
\end{figure}

From the comparison of the fit parameters for data and AMPT, we
observe that the AMPT can be used as a proxy for experimental data. 
The AMPT points are 
fitted with power law fits, shown in Fig.~\ref{param} as dashed lines. 
These fit values provide a way to compute the \Eta-distribution at 
any collision energy and centrality. Accordingly, 
these fit values are extended up to higher energy,
{\it viz.}, \sNN~=~5.02~TeV. 
The Gaussian fit parameters, along with the extrapolated 
values for \sNN~=~5.02~TeV from AMPT are presented in Table~\ref{table1}. 
With the extrapolated parameter set for Pb+Pb collisions at 
\sNN~=~5.02~TeV, the \Eta-distributions at different 
collision energies are obtained. The results are shown in
Fig.~\ref{eta5020}.

\section{Energy dependence of global parameters}

Parameterization of \Eta-distributions of charged particles 
from the AMPT model can be used 
to obtain energy dependence of several other global observables.
Here we discuss the collision energy dependence of charged particle multiplicity
density at mid-rapidity, centrality dependence of charged particle
multiplicity density and the collision energy
dependence of Bjorken energy density.

\begin{figure}[tbp]
\centering \includegraphics[width=0.49\textwidth]{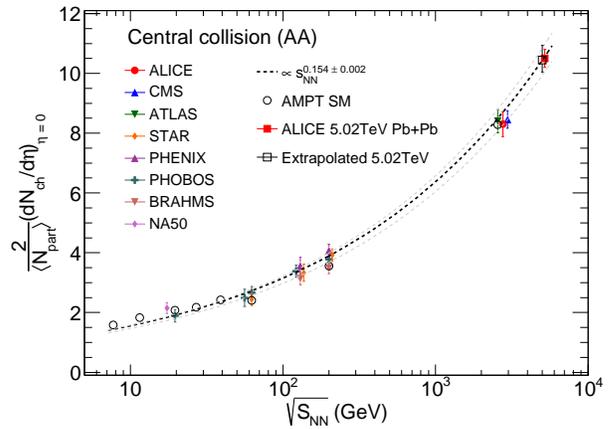} 
\caption{
Pseudorapidity density of charged particles, normalized to
number of participant pairs ($\frac{2}{\langle N_{\rm part} \rangle}(dN_{\rm ch}/d\eta)$),
plotted as a function of 
collision energy for central Au+Au or Pb+Pb collisions from
experimental data and AMPT model. 
Some of the data points are shifted along $x$-axis for clarity of presentation.  
}
\label{rapdens}
\end{figure}

The quantity, $2(dN_{\rm ch}/d\eta)$/\aveNpart, gives the
charged particle multiplicity density at $\eta$=0 scaled
by the average number of participant pairs (\aveNpart/2). 
Figure~\ref{rapdens} shows the variation of this quantity as a
function of \sNN~for central (top 5\% cross section) collisions. The
plot shows an increase in the multiplicity density with the increase
of the collision energy. The data points are taken from PHOBOS,
BRAHMS, STAR,
and PHENIX experiments at RHIC and ALICE, CMS and ATLAS  experiment at
LHC. The results from AMPT model are shown by solid red points. For  
Pb+Pb data at 5.02~TeV, the extrapolated results from  
Fig.~\ref{eta5020} have been plotted.  The AMPT
results explain the data quite well. 
A power law fit to the AMPT model
data gives the fit value as (0.77 $\pm$ 0.04)$ s_{\rm   {NN}}^{0.154 \pm 0.002}$.
 This matches the fit given in Ref.~\cite{alice5}. As shown in
the figure, the extrapolated
value at \sNN~=~5.02~TeV is close to the recently published data from
the ALICE experiment~\cite{alice5}. The beam energy dependence of
charged particle multiplicity density has been studied for other
centralities.
Power law fit to each of the curves give the $s_{\rm NN}$ dependence 
as $s_{\rm NN}^{0.154}$ to  $s_{\rm NN}^{0.109}$ from top central (0-5\%) 
to peripheral (70-80\%) collisions. This is consistent with the
conclusion that the particle multiplicity increases faster for central
collisions compared to peripheral collisions.  

\begin{figure}[tbp]
\centering \includegraphics[width=0.49\textwidth]{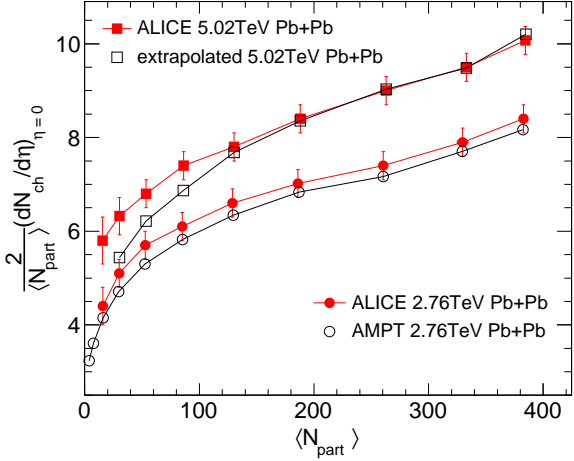} 
\caption{Centrality dependence of $\frac{2}{\langle N_{\rm part}\rangle}(dN_{\rm ch}/d\eta)$ for
  Pb+Pb collisions at \sNN~=~2.76 and 5.02~TeV. AMPT model calculations
  for \sNN~=~2.76~TeV and extrapolations for \sNN~=~5.02~TeV reasonably
  explain the ALICE data~\cite{alice2,alice5}.
}
\label{5020}
\end{figure}

The centrality dependences of charged particle multiplicity density
have been reported for Pb+Pb collisions at \sNN~=~2.76~TeV~\cite{alice2} and
5.02~TeV~\cite{alice5}. As discussed earlier, the AMPT model
calculations describe the data well at \sNN~=~2.76~TeV.
By extrapolating the fit parameters from the AMPT model to higher energies of
\sNN~=~5.02~TeV, we obtain the centrality dependence of charged
particle multiplicity density at this energy. For central (0-5\%)
collisions, the multiplicity density comes out to be $1964\pm 30$.
The results from the
experimental data and AMPT calculations for both \sNN~=~2.76 and
5.02~TeV as a function of centrality are shown in Fig.~\ref{5020}.  For Pb+Pb collisions at 
\sNN~=~2.76~TeV, the AMPT results are within the experimental errors. 
For Pb+Pb collisions at \sNN~=~5.02~TeV, the AMPT results agree with
the experimental data points, except for peripheral collisions with
\aveNpart~less than 130.

\begin{figure}[tbp]
\centering \includegraphics[width=0.49\textwidth]{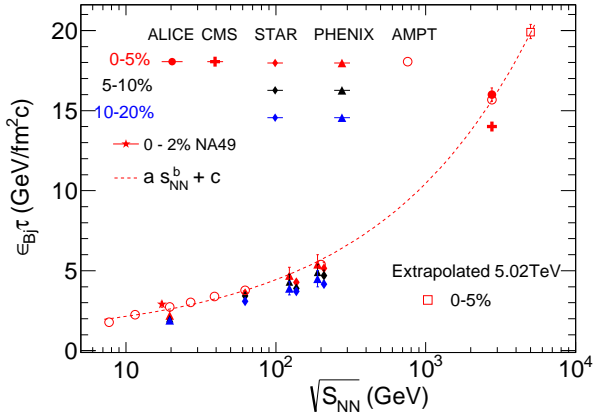} 
\caption{
Energy density (\Ebj) as a function of \sNN~for
experimental 
data~\cite{na49_et,star_et1,star_et2,phenix_et1,phenix_et2,phenix_et3,alice_et1,alice_et2,cms_et}
 and AMPT model. Power law fits to the AMPT results
 are extrapolated to
\sNN~=~5.02~TeV. Some of the data points are shifted along $x$-axis for clarity of presentation.  
}
\label{density}
\end{figure}

Charged particle multiplicity density is normally used to estimate the
initial energy density of the fireball by using 
the Bjorken estimation given as~\cite{bjorken}:
\begin{eqnarray}
\epsilon_{\rm Bj} = \frac{1}{\pi R^2 \tau} \frac{dE_{\rm T}}{dy},
\end{eqnarray}
where $\tau$ is the formation time, $\pi R^2$ is the effective 
area of the fireball or the overlap area of the colliding nuclei, and
$dE_{\rm T}$ is the total initial energy within a rapidity 
window $dy$. The last term can be approximated as~\cite{star_et2}:
\begin{eqnarray}
\frac{dE_{\rm T}}{dy} \approx
\frac{3}{2} \bigg(\langle m_{\rm T} \rangle \frac{dN}{dy}\bigg)_{\pi^{\pm}} +
2 \bigg(\langle m_{\rm T} \rangle \frac{dN}{dy}\bigg)_{K^{\pm},p,\bar{p}}.
\end{eqnarray}
$\langle m_{\rm T} \rangle$ is the mean transverse mass of identified particles
($\pi^{\pm}$, $K^{\pm}$, $p$ or $\bar{p}$).
The value of $\tau$ is typically taken as 1~fm. But 
in the absence of experimental knowledge of $\tau$, the
energy density is expressed in terms of \Ebj.

The energy density, \Ebj, as a function of collision energy is
presented in Fig.~\ref{density} for experimental results at
three centralities from
NA49~\cite{na49_et}, STAR~\cite{star_et1,star_et2}, 
PHENIX~\cite{phenix_et1,phenix_et2,phenix_et3},
ALICE~\cite{alice_et1,alice_et2} and CMS~\cite{cms_et}
collaborations. 
In some cases, there are differences in experiments results at same
collision energies show different results. 
AMPT model results are superimposed for central (0-5\%) collisions.
It is observed that the AMPT results reasonably describe the
experimental data.
The AMPT results of \Ebj~are fitted with a power law (for central
$\propto$ $s^{0.22\pm0.015}$) for different
centralities. 
For central (0-5\%) collisions, the value of \Ebj~comes out to be 19.88 $\pm$ 0.48~GeV/fm$^2$c.
The value of the exponent in the power law fits are observed to vary from
$s_{\rm NN}^{0.22}$ to  $s_{\rm NN}^{0.10}$ for central (0-5\%) to
peripheral (70-80\%) collisions, respectively.
\Ebj~is a combination
of $dN_{\rm ch}/d\eta$ and $\langle m_{\rm T} \rangle$~, both of which vary as 
power law with respect to collision energy. That may explain the origin
of the power law behavior of energy density. As a function of
collision energy, the energy density increases much faster for central
collisions compared to peripheral collisions.


\section{Summary}

We have studied the \Eta-distributions of produced charged particles
for Au+Au collisions at \sNN~=~7.7 to 200~GeV, corresponding to the
collisions at RHIC and for Pb+Pb collisions at \sNN~=~2.76~TeV,
corresponding to the collisions at LHC. We have employed the string
melting mode of the AMPT model to describe the experimental data. We
observe that using the total parton elastic cross section,
\sgg~=~10~mb,
the AMPT model can 
explain the RHIC data, whereas \sgg~=~1.5~mb is needed for explaining the
data at LHC.  AMPT model, with these settings are used to further
study the \Eta-distributions and initial energy densities. 
The shapes of the \Eta-distributions could be
explained by using double Gaussian functions with a set of parameters
comprising of the amplitude, the position of the peaks in $\eta$,
and the widths of the distributions. As expected, 
with the increase of the beam energy, the amplitudes increase,
the peak positions move farther apart, and the widths of the
distributions increase. The parameters are fitted well by power law
fits, using which the pseudorapidity distributions can be obtained for
any beam energy and collision centrality. 
We obtain initial energy density as a function of collision energy and
collision centrality using Bjorken formalism.
Power law fits to the 
multiplicity density at mid-rapidity give the $s_{\rm NN}$ dependence 
as $s_{\rm NN}^{0.154}$ to  $s_{\rm NN}^{0.109}$ from top central (0-5\%) 
to peripheral (70-80\%) collisions. Similarly, power law fits to the
energy density yield the $s_{\rm NN}$ dependence as
$s_{\rm NN}^{0.22}$ to  $s_{\rm NN}^{0.10}$ for the same centrality
ranges.  As a function of collision energy, the particle multiplicity 
and energy density increase much faster for central collisions compared to the 
peripheral collisions. Extrapolating the parameters 
to collisions at \sNN~=~5.02~TeV, we are able to explain the 
recently published results on centrality dependence of charged particle 
multiplicity and energy density. At this energy, the pseudorapidity density of charged
particles for central (0-5\%) collisions is $1964\pm30$ and energy
density, \Ebj~is 19.98~GeV/fm$^2$c.
Furthermore, we note that the results obtained in the present study 
can be interpolated for intermediate energies to obtain
\Eta-distributions and energy densities for heavy-ion
collisions in the Facility for Antiproton and Ion Research (FAIR). 
For laboratory energy of 11~GeV at FAIR, the energy density would be 1.8
GeV/fm$^3$ for $\tau=1$~fm, which is an interesting region to study
the deconfined matter at high net-baryon density.

\medskip

\noindent
{\bf Acknowledgement}
This research used resources of the  LHC grid computing centre at the Variable Energy Cyclotron Centre.

\end{document}